\documentclass{article}
\begin{document}
\newcommand{\R}{\mathbb{R}}
\newcommand{\Z}{\mathbb{Z}}
\newcommand{\N}{\mathbb{N}}
\newcommand{\Q}{\mathbb{Q}}

\newcommand{\bcal}{\mathcal{B}}

\title{The Future of Search and Discovery in Big Data Analytics: 
Ultrametric Information Spaces}

\author{F. Murtagh (1, 2) and P. Contreras (2, 3) \\
(1) Science Foundation Ireland \\
(2) Thinking Safe Ltd. \\
(3) Computer Learning Research Centre \\
Department of Computer Science \\ 
Royal Holloway, University of London 
}

\maketitle

\begin{abstract}
Consider observation data, comprised of $n$ observation
vectors with values on a set of attributes.  This gives 
us $n$ points in attribute space.   
Having data structured as a tree, implied by having our 
observations embedded in an ultrametric topology, offers 
great advantage for proximity searching.  If we have 
preprocessed data through such an embedding, then 
an observation's nearest neighbor is found in constant
computational time, i.e.\ $O(1)$ time.   A 
further  powerful approach is discussed
in this work: the inducing of a hierarchy, and hence
a tree, in linear computational time, i.e.\ $O(n)$ time 
for $n$ observations.   It is with such a basis for 
proximity search and best match that we can address 
the burgeoning problems of processing very large,
and possibly also very high dimensional,  data sets.  
\end{abstract}

\section{Introduction}

Under the heading of ``Addressing the big data challenge'', the European
7th Framework Programme sees the issue thus (see INFSO, 2012): 
``Recent industry reports detail how data volumes are growing at a 
faster rate than our ability to interpret and exploit them for 
innovative ICT applications, for decision support, planning, 
monitoring, control and interaction. This includes unstructured data 
types such as video, audio, images and free text as well as structured 
data types such as database records, sensor readings and 3D. While each 
of these types requires some specific form of
processing and analytics, many of the general principles for managing 
and storing them at extreme scales are common across all of them.''
Analytics tool capability is called for, to address these burgeoning 
issues in the data intensive industries, to support ``effective policy 
making and implementation'' of public bodies resulting in ``significant
annual savings from Big Data applications'', and also to exploit open, 
linked data -- ``foster the reuse of public sector information and 
strengthen other open data activities linked to commercial exploitation.''
The ``big data'' marketplace is stated to be potentially worth 
approximately USD 600 billion. 

To address the challenges of search and discovery in massive 
and complex data sets and data flows, 
it is our contention in this work that we must move to an appropriate 
topology -- to an appropriate framework such that computation is greatly 
facilitated.  Our work is all about empowering those who are involved 
in data analytics, through clustering and related 
algorithms, to face these new challenges.  Scalability and interactivity 
are two of the performance issues that follow directly from 
clustering algorithms, for search, retrieval and discovery, 
that are of linear computational complexity or better (logarithmic, or 
constant).   

\section{Ultrametric Information Spaces}

For high dimensional spaces and also for massive data spaces, it has
been shown in Murtagh (2004) that one can exploit both symmetry and 
sparsity to great effect in order to carry out nearest neighbor 
or best match search and other related operations.  

The triangular inequality holds for a metric space: 
$d(x,z) \leq d(x,y) + d(y,z)$ for any triplet of points, $x, y, z$. 
In addition 
the properties of symmetry and positive definiteness are respected. The 
``strong triangular inequality'' or ultrametric inequality is:
$d(x,z) \leq \mbox{max} \{d(x,y), d(y,z)\}$ 
for any triplet $x, y, z$. An ultrametric space 
(Benz\'ecri, 1979; van Rooij, 1978) implies respect for a 
range of stringent properties. For example, the triangle formed by any 
triplet is necessarily isosceles, with the
two large sides equal; or is equilateral.

\subsection{Computational Costs of Operations in an Ultrametric Space}

Given that sparse forms of coding are considered for how complex stimuli 
are represented in the cortex (see Young and Yamane, 1992), the 
ultrametricity of such spaces becomes important because of this 
sparseness of coding. Among other implications, this points to the 
possibility that semantic pattern matching is best accomplished 
through ultrametric computation.

A convenient data structure for points in an ultrametric space 
is a dendrogram. We define a dendrogram as a rooted, labeled, 
ranked, binary tree (Murtagh, 1984a). For $n$ observations,
with such a definition of tree, there 
are precisely $n - 1$ levels. With each level there is an associated 
rank $1, 2, \dots , n - 1$, with level 1 corresponding to the singletons,
and level $n - 1$ corresponding to the root node, and also to the
cluster that encompasses all observations.  With such a tree, there is 
an associated distance on the tree, termed the ultrametric distance, which 
is a mapping (of the Cartesian product of the observation set with 
itself) into the positive reals.  

We will use the terms point and observation interchangeably, when 
the context allows.  That is to say, an observation vector is a point 
in a space of ambient dimensionality defined by the cardinality of the
attribute set, on which the observation takes values.   

Operations on binary trees are often based on tree traversal between 
root and terminal. See e.g.\ van Rijsbergen (1979).  
Hence computational cost of such operations is 
dependent on root-to-terminal(s) path length. The total path length 
of a root-to-terminal traversal varies for each terminal (or point 
in the corresponding ultrametric space). It is  simplest to 
consider path length in terms of level or tree node rank (and if it 
is necessary to avail of path length in terms of ultrametric distances, 
then constant computational time, only, is needed for table lookup). 
A dendrogram's root-to-terminal path length can vary from close to 
log$_2 n$ (``close to'' because the path length has to be an 
integer) to $n - 1$ (Murtagh, 1984b). Let us call this computational 
cost of a tree traversal $O(t)$.

Most operations that we will now consider make use of a dendrogram 
data structure. Hence the cost of building a dendrogram is important. 
For the problem in general, see K\v{r}iv\'anek  and Mor\'avek (1984, 1986)
 and Day (1996). For $O(n^2)$ implementations of most commonly used 
hierarchical clustering algorithms, see Murtagh (1983, 1985).
In section \ref{sect2} we will address the issue of efficiently 
constructing a hierarchical clustering, and hence mapping observed
data into an ultrametric topology.  We will discuss a linear time 
approach for this.   

To place a new point (from an ultrametric space) into a dendrogram, we 
need to find its nearest neighbor. We can do this, in order to write 
the new terminal into the dendrogram, using a root-to-terminal traversal 
in the current version of a dendrogram. This leads to our first proposition.

\medskip
\noindent
{\bf Proposition 1}: The computational complexity of adding a new terminal 
to a dendrogram is $O(t)$, where $t$ is one traversal from root to 
terminals in the dendrogram.

\medskip
\noindent
{\bf Proposition 2}: The computational complexity of finding the 
ultrametric distance between two terminal nodes is twice the length of 
a traversal from root to terminals in the dendrogram. Therefore distance 
is computed in $O(t)$ time.
Informally: we potentially have to traverse from each terminal to the root 
in order to find the common, ``parent'' node.

\medskip
\noindent
{\bf Proposition 3}: The traversal length from dendrogram root to 
dendrogram terminals is best case 1, and worst case $n - 1$. When the 
dendrogram is optimally balanced or structured, the traversal length 
from root to terminals is $\lfloor log_2 n \rfloor$, where $\lfloor . 
\rfloor$ is the floor, or integer part, function. Hence $1 \geq O(t) 
\geq  n - 1$, and for a balanced tree $O(t) = \log_2 n$.

Depending on the agglomerative criterion used, we can approximate 
the balanced or structured dendrogram -- and hence favorable case -- 
quite well in practice (Murtagh, 1984b).  The Ward, or minimum variance,
agglomerative criterion is shown empirically to be best.   

\medskip
\noindent
{\bf Proposition 4}: Nearest neighbor search in ultrametric space 
can be carried out in O(1) or constant time.

\medskip

This results from the following: the nearest neighbor pair must be 
in the same tightest cluster that contains them both. There is only 
one candidate to check for in a dendrogram. Hence nearest neighbor 
finding results in firstly finding the lowest level cluster containing 
the given terminal; followed by finding the other terminal in this 
cluster. Two operations are therefore required.

\subsection{Implications}

In Murtagh (2004a, 2004b) we 
have shown that high dimensional and sparse codings tend to be 
ultrametric. This is an interesting result in its own right. However 
a far more important result is that certain computational operations 
can be carried out very efficiently indeed in space endowed with an 
ultrametric.

Chief among these computational operations, we have noted, is that 
nearest neighbor finding can be carried out in (worst case) constant 
computational time, relative to the number of observables considered, $n$. 
Depending on the structure of the ultrametric 
space (i.e. if we can build a balanced dendrogram data structure), 
pairwise distance calculation can be carried out in logarithmic 
computational time.

We have also (Murtagh, 2004a)
reviewed approaches to using ultrametric distances in 
order to expedite best match, or nearest neighbor, or more generally 
proximity search. The usual constructive approach, viz.\ build a 
hierarchic clustering, is simply not computationally feasible in 
very high dimensional spaces as are typically
found in such fields as speech processing, information retrieval, 
or genomics and proteomics.

Forms of sparse coding are considered to be used 
in the human or animal cortex. We raise the interesting question as 
to whether human or animal thinking can be computationally efficient 
precisely because such computation is carried out in an ultrametric space.
For further elaboration on this, see Murtagh (2012a, 2012b).  

\section{Linear Time and Direct Reading Hierarchical Clustering}
\label{sect2}

In areas such as search, matching, retrieval and general data analysis,
massive increase in data requires new methods that can cope well
with the explosion in volume and dimensionality of the available data.
The Baire metric, which is furthermore an ultrametric, has particular
advantages when used
to induce a hierarchy and in turn to support clustering, matching and other 
operations.  See Murtagh and Contreras (2012), and Contreras and Murtagh
(2012).  

Arising directly out of the Baire distance is an ultrametric tree, which 
also can be seen as a tree that hierarchically clusters data. This  
presents a number of advantages when storing and retrieving data. When 
the data source is in numerical form this ultrametric tree can be used 
as an index structure making matching and search, and thus retrieval, 
much easier.

The clusters can be associated with hash keys, that is to say, the 
cluster members can be mapped onto ``bins'' or ``buckets''.

Another vantage point in this work is precision of measurement.  Data 
measurement precision can be
either used as given or modified in order to enhance the inherent ultrametric
and hence hierarchical properties of the data.

Rather than mapping pairwise relationships onto the reals, as distance does, we
can alternatively map onto subsets of the power set of, say, attributes of our
observation set.  This is expressed by the generalized ultrametric, which
maps pairwise relationships into a partially ordered set (see 
Murtagh, 2011).  It is also current 
practice
as formal concept analysis where the range of the mapping is a lattice.

Relative to other algorithms the Baire-based hierarchical clustering method is fast.
It is a direct reading algorithm involving one scan of the input data set, and
is of linear computational complexity.

Many vantage points are possible, all in the Baire metric framework.  The 
following
vantage points are discussed in Murtagh and Contreras (2012).

\begin{itemize}
\item Metric that is simultaneously an ultrametric.
\item Hierarchy induced through m-adic encoding (m positive integer, 
e.g.\ 10).
\item p-Adic (p prime) or m-adic clustering.
\item Hashing of data into bins.
\item Data precision of measurement implies
how hierarchical the data is.
\item Generalized ultrametric.
\item Lattice-based formal concept analysis.
\item Linear computational time hierarchical clustering.
\end{itemize}

\subsection{Ultrametric Baire Space and Distance}
\label{subsection:baire}

A Baire space consists of countably infinite sequences with a metric defined
in terms of the longest common prefix: the longer the common prefix, the
closer a pair of sequences. What is of interest to us is this longest
common 
prefix metric, which we call the Baire 
distance (Bradley, 2009; Mirkin and Fishburn, 1979; Murtagh et al., 2008).

We begin with the
longest common prefixes at issue being digits of precision of
univariate or scalar values.
 For example, let us consider two such decimal
values, $x$ and $y$, with both measured to some maximum precision.
We take as maximum precision the length of the value with the
fewer decimal digits.
With no loss of generality we take $x$ and $y$ to be bounded by 0 and 1.
Thus we consider ordered sets $x_{k}$ and $y_{k}$ for $k \in K$.
So $k = 1$ is the first decimal place of precision;
$k = 2$ is the second decimal place; . . . ; $k = \left| K \right|$ is
the $\left| K \right|th$ 
decimal place.  The cardinality of the set K is the precision with
which a number, $x$ or $y$, is measured.

Consider as examples $x_{3} = 0.478$; and $y_{3} = 0.472$. Start from the
first decimal position.
For $k = 1$, we find $x_{1} = y_{1} = 4$. For $k = 2$, $x_{2} = y_{2} = 7$.
But for $k = 3$,  $x_{3} \neq y_{3}$.

We now introduce the following distance (case of vectors $x$ and $y$, with
1 attribute, hence unidimensional):

\begin{equation}
\label{eq:baire}
{\rm d}_\bcal(x_{K}, y_{K}) =
        \left\{
        \begin{array}{ll}
       1 &\;\; $if$\;\;  x_{1} \neq y_{1}\\
       $inf$\;\;  \bcal^{-\nu} & \;\;\;\;\;\; x_{\nu} = y_{\nu}, \;\;\; 1 \leq \nu \leq
       \left| K \right|
    \end{array}
    \right.
\end{equation}

We call this ${\rm d}_\bcal$ value Baire distance, which is a 1-bounded
ultrametric (Bradley, 2009; Murtagh, 2007) distance, $0 < {\rm d}_\bcal \leq 1$.
When dealing with binary (boolean)
data 2 is the chosen base, $\bcal = 2$. When working with real
numbers the base is best defined to be 10, $\bcal = 10$.
With $\bcal = 10$, for instance, it can be seen that the Baire
distance is embedded in a 10-way tree which leads to a convenient
data structure to support search and other operations when we have
decimal data.  As a consequence data can be organized, stored and
accessed very efficiently and effectively in such a tree.

For $\bcal$ prime, this distance has been studied by Benois-Pineau
et al.\ (2001)
and by Bradley (2009, 2010), with many further (topological
and number theoretic, leading to algorithmic and computational)
insights arising from the
p-adic (where p is prime) framework. See also Anashin and Khrennikov
(2009).

For use of random projections to allow for analysis of 
multidimensional data in the 
scope of the Baire distance, see Contreras and Murtagh (2012) and 
also Murtagh and Contreras (2012).  In these works, a range of 
very large data sets are considered, for clustering and for proximity 
search, in domains that include astronomy (photometric and 
astrometric redshifts), and chemoinformatics.

\subsection{Linear Time, or $O(N)$ Computational Complexity, Hierarchical 
Clustering}
\label{subsect81}

A point of departure for our work has been the computational objective
of bypassing computationally demanding hierarchical clustering
methods (typically quadratic time, or $O(n^2)$ for $n$ input
observation vectors), but also having a framework that is
of great practical importance in terms of the application domains.

Agglomerative hierarchical clustering algorithms are based on
pairwise distances (or dissimilarities) implying computational
time that is $O(n^2)$ where $n$ is the number of observations.
The implementation required to achieve this is, for most agglomerative
criteria, the nearest neighbor chain, together with the reciprocal
nearest neighbors, algorithm
(furnishing inversion-free
hierarchies whenever Bruynooghe's reducibility property, see 
Murtagh (1985), 
is satisfied by the cluster criterion).

This quadratic time requirement is a worst case performance result.
It is most often the average time also since the pairwise agglomerative
algorithm is applied directly to the data without any preprocessing
speed-ups (such as preprocessing that facilitates fast nearest neighbor
finding).   An example of a linear average time algorithm for
(worst case quadratic computational time) agglomerative hierarchical
clustering is in Murtagh (1983).  

With the Baire-based hierarchical clustering algorithm, we have an algorithm
 for linear time worst case hierarchical clustering.  It can be characterized
as a divisive rather than an agglomerative algorithm.

\subsection{Grid-Based Clustering Algorithms}
\label{subsection:grid-based-clustering}

The Baire-based hierarchical clustering algorithm has characteristics
that are related to grid-based clustering algorithms, and density-based 
clustering
algorithms, which -- often -- were developed in order to handle very large
data sets.

The main idea here is to use a grid like structure to split the information 
space, separating the dense grid regions from the less dense ones to form 
groups.
In general, a typical approach within this category will consist of 
the following steps (Grabusts and Borisov, 2002):
\begin{enumerate}
        \item Creating a grid structure, i.e.\ partitioning
    the data space into a finite number of  non-overlapping cells.
        \item Calculating the cell density for each cell.
        \item Sorting of the cells according to their densities.
        \item Identifying cluster centers.
        \item Traversal of neighbor cells.
\end{enumerate}

Additional background on grid-based clustering can be
found in the following works: Chang and Jin (2002), 
Gan et al.\ (2007), Park and Lee (2004), and  Xu and Wunsch (2008).

Cluster bins, derived from an m-adic tree, provide us with
a grid-based framework or data structuring.  We can read off the
cluster bin members from an m-adic tree.  
An m-adic tree requires one
scan through the data, and therefore this data structure is
constructed in linear computational time.

In such a preprocessing context, clustering with the Baire distance
can be seen as a ``crude'' method
for getting clusters.
After this we can use more traditional techniques
to refine the clusters in terms of their membership.
Alternatively (and we have quite extensively compared Baire clustering
with, e.g.\
k-means, where it compares very well, see Murtagh et al., 2008, and 
Contreras and Murtagh, 2012) clustering with the Baire distance
can be seen as fully on a par with any optimization algorithm for
clustering.  As optimization, and just as one example from the many
examples reviewed in this article, the Baire approach optimizes
an m-adic fit of the data simply by reading the m-adic structure
directly from the data.

\section{Conclusions}

Baire distance is an ultrametric, so we can think of reading off
observations as a tree.

Through data precision of measurement, alone, we can enhance inherent
ultrametricity, or inherent hierarchical properties in the data.

 Clusters in such a Baire-based hierarchy are simple ``bins'' and
assignments are determined through a very simple hashing.   (E.g.\
$0.3475 \longrightarrow$ bin 3, and $\longrightarrow$ bin 34, and
$\longrightarrow$ bin 347, and $\longrightarrow$ bin 3475.)

As we have observed, certain 
search-related computational operations can be carried out 
very efficiently indeed in space endowed with an ultrametric.
Chief among these computational operations is that 
nearest neighbor finding can be carried out in (worst case) constant 
computational time. Depending on the structure of the ultrametric 
space (i.e.\ if we can build a balanced dendrogram data structure), 
pairwise distance calculation can be carried out in logarithmic 
computational time.

In conclusion we have here a comprehensive approach, founded on 
ultrametric topology rather than more traditional metric geometry, 
in order to address the burgeoning problems presented by ``big data''
analytics, i.e.\ massive data sets in potentially very high dimensional
spaces.

\section*{References}

\begin{enumerate}
\item 
V. Anashin and A. Khrennikov, {\em 
Applied Algebraic Dynamics}, De Gruyter, 2009.

\item 
J. Benois-Pineau, A.Yu. Khrennikov and N.V. Kotovich, ``Segmentation
of images in p-adic and Euclidean metrics'', {\em Dokl.\ Math.},
64, 450--455, 2001.  

\item J.P. Benz\'ecri, {\em  La Taxinomie}, Dunod, 2nd edition, 1979.

\item P.E. Bradley, ``On p-adic classification'', {\em 
p-Adic Numbers, 
Ultrametric Analysis, and Applications}, 1:271--285, 2009. 

\item P.E. Bradley, ``Mumford dendrograms'', Journal of Classification, 
53:393--404, 2010.

\item Jae-Woo Chang and Du-Seok Jin, ``A new cell-based clustering method for 
large, high-dimensional data in data mining applications'', in {\em 
SAC '02: 
Proceedings of the 2002 ACM symposium on Applied
computing}, pages 503--507, 2002. 

\item 
P. Contreras and F. Murtagh, ``A very fast, linear time p-adic and
hierarchical clustering algorithm using the Baire metric'', 
{\em Journal of
Classification}, forthcoming, 2012.

\item 
W.H.E. Day, ``Complexity theory: An introduction for 
practitioners of classification'', in P. Arabie, L.J. Hubert and 
G. De Soete, Eds., {\em Clustering and Classification}, Singapore:
World Scientific, 199--233, 1996.

\item Guojun Gan, Chaoqun Ma, and Jianhong Wu,
{\em Data Clustering Theory, 
Algorithms, and Applications}, Society for Industrial and Applied 
Mathematics. SIAM, 2007.

\item P. Grabusts and A. Borisov, ``Using grid-clustering methods in data 
classification'', In {\em 
PARELEC ’02: Proceedings of the International 
Conference on Parallel Computing in Electrical Engineering}, p. 
425, Washington, DC, USA, 2002. IEEE Computer Society.

\item INFSO -- Directorate General Information Society, European Commission,
{\em FP7 ICT Work Programme 2013 Orientations, Overview}, Version 
10/01/2012, white paper INFSO-C2/10/01/2012.  

\item M. K\v{r}iv\'anek and J. Mor\'avek,  
``NP-hard problems in hierarchical-tree clustering'', {\em Acta 
Informatica}, 23, 311--323, 1986.
 
\item M. K\v{r}iv\'anek and J. Mor\'avek, 
``On NP-hardness in hierarchical clustering'',
in T. Havr\'anek, Z. Sid\'ak and M. Nov\'ak, ed., 
{\em Compstat 1984: Proceedings in Computational Statistics}, 
189--194, Vienna: Physica-Verlag, 1984.

\item B. Mirkin and P. Fishburn, {\em Group Choice}, 
V.H. Winston, 1979.  

\item F. Murtagh, ``Expected time complexity results for hierarchic 
clustering algorithms that use cluster
centers'', {\em Information Processing Letters}, 16:237--241, 1983. 

\item F. Murtagh, {\em Multidimensional Clustering Algorithms}, 
Physica-Verlag, 1985.

\item F. Murtagh, ``Counting dendrograms: a survey'', 
{\em Discrete Applied Mathematics}, 7, 191--199, 1984a.

\item F. Murtagh, ``Structures of hierarchic clusterings: 
Implications for information retrieval and for multivariate 
data analysis'', {\em Information Processing and Management}, 
20, 611--617, 1984b.

\item F. Murtagh, ``On ultrametricity, data coding, and computation'',
{\em Journal of Classification}, 21, 167--184, 2004a.

\item 
F. Murtagh, ``Thinking ultrametrically'', in D. Banks, 
L. House, F.R. McMorris, P. Arabie and W. Gaul, Eds., 
{\em Classification, Clustering, and Data Mining Applications}, 
Springer, 3--14, 2004b.

\item 
F. Murtagh, G. Downs, and P. Contreras, ``Hierarchical clustering of massive,
high dimensional datasets by exploiting ultrametric embedding'', 
{\em SIAM 
Journal on Scientific Computing}, 30(2):707-730, February
2008.

\item F. Murtagh, ``Ultrametric and generalized ultrametric 
in logic and in data analysis'', in T.S. Clary, Ed., 
{\em Horizons in Computer Science, Volume 2}, pp. 251-267, Nova Science
Publishers, 2011.  

\item F. Murtagh and P. Contreras, ``Fast, linear time, m-adic 
hierarchical clustering for search and retrieval using the 
Baire metric, with linkages to generalized ultrametrics, hashing,      
formal concept analysis, and precision of data measurement'',
{\em p-Adic Numbers,
Ultrametric Analysis, and Applications}, 4, 45--56, 2012.

\item F. Murtagh, ``Ultrametric model of mind, I: Review'',  \\
http://arxiv.org/abs/1201.2711, 
2012a.

\item F. Murtagh, ``Ultrametric model of mind, II: Application to 
text content analysis'', http://arxiv.org/abs/1201.2719, 
2012b.

\item Nam Hun Park and Won Suk Lee,
``Statistical grid-based clustering over
datastreams'', {\em SIGMOD Record}, 33(1):32--37, 2004.

\item 
C.J. van Rijsbergen, {\em Information Retrieval}, Butterworths, 
1979.

\item 
A.C.M. van Rooij,
{\em Non-Archimedean Functional Analysis}. Marcel Dekker, 1978.

\item
M.P. Young and S. Yamane, ``Sparse population coding of 
faces in the inferotemporal cortex'', {\em Science}, 256, 
1327--1331, 1992.

\item
Rui Xu and D.C. Wunsch, {\em Clustering}, IEEE Computer Society Press, 2008.

\end{enumerate}

\end{document}